# Knowledge-aided Two-dimensional Autofocus for Spotlight SAR Polar Format Imagery[1]

Xinhua Mao

College of Electronic and Information Engineering, Nanjing University of Aeronautics and Astronautics, Nanjing, Jiangsu 210016, China.   Email: xinhua@nuaa.edu.cn

**ABSTRACT**

Conventional two-dimensional (2-D) autofocus algorithms blindly estimate the phase error in the sense that they do not exploit any a priori information on the structure of the 2-D phase error. As such, they often suffer from low computational efficiency and lack of data redundancy to accurately estimate the 2-D phase error. In this paper, a knowledge-aided (KA) 2-D autofocus algorithm which is based on exploiting a priori knowledge about the 2-D phase error structure, is presented. First, as a prerequisite of the proposed KA method, the analytical structure of residual 2-D phase error in SAR imagery is investigated in the polar format algorithm (PFA) framework. Then, by incorporating this a priori information, a novel 2-D autofocus approach is proposed. The new method only requires an estimate of azimuth phase error and/or residual range cell migration, while the 2-D phase error can then be computed directly from the estimated azimuth phase error or residual range cell migration. This 2-D autofocus method can also be applied to refocus moving targets in PFA imagery. Experimental results clearly demonstrate the effectiveness and robustness of the proposed method.

## I. Introduction

Synthetic aperture radar (SAR) is an imaging system which coherently processes multiple echo pulses to generate scene image with a high azimuth resolution. Coherent data processing requires accurate

---

estimation of the relative geometric relationship between the radar's flight path and the scene being imaged. Such geometry information is typically obtained from motion sensors, such as inertial measurement units (IMUs) and global positioning system (GPS) receivers. These sensors, however, could be too expensive or may not provide the accuracy required for ultra-high resolution imaging. In addition, signal propagation through turbulent media is another critical factor that prohibits a SAR from achieving a high image resolution [2, 3]. Consequently, signal based motion compensation, i.e., autofocus, is often an indispensable step that provides a necessary supplement to IMU/GPS devices, especially for airborne SAR processing with a very fine resolution.

Range errors caused by measurement and/or atmosphere perturbations impose two adverse effects on the echoes, namely, azimuth phase error (APE) and residual range cell migration (RCM). The APE makes an image defocused in the azimuth dimension, whereas the residual RCM introduces two-dimensional (2-D) defocusing in both azimuth and range dimensions. There exists a simple linear relationship between these two types of errors in the phase history domain, i.e., the APE is the product of the residual RCM and a constant factor of $4\pi/\lambda$ [4, 5], where $\lambda$ is the wavelength. However, this linear relationship no longer holds after image formation processing [5-8]. In addition, high-order terms in the range frequency may be introduced in the image formation process, causing re-defocusing in the range direction. For a SAR system operated in a submeter wavelength ($4\pi/\lambda \gg 1$), the APE has a much more significant impact on the formed image than the residual RCM. When the range error is relatively small, e.g., within a range resolution cell, the effect of the residual RCM can be neglected and, as such, only the APE should be compensated. This is the general presumption made in most existing autofocus algorithms, such as Mapdrift (MD) [9], Phase Difference (PD) [10], Phase Gradient Autofocus (PGA) [11], Eigenvector Method [12], and Maximum-likelihood Autofocus (MLA) [13]. As the SAR resolution becomes finer, however, the increased size of synthetic aperture makes the accumulated range error more pronounced. On the other hand, range resolution cell becomes smaller. As a result, the residual RCM will inevitably exceed the range resolution cell size in ultra-high resolution SAR systems [5, 6]. In this situation, 2-D autofocus becomes a necessary procedure to obtain refocused images.

Two alternative strategies are available for the estimation of 2-D phase error. The first one is to estimate the error in a blind manner as they assume that the 2-D phase error is absolutely unknown a priori [14-18]. These approaches face two major hurdles if they are to become as useful as one-dimensional (1-D) autofocus. First, compared with 1-D phase error estimation, the data have much less available redundancy for 2-D phase error estimation. Second, the increased dimensions of the estimated parameters result in high computational complexity, thereby limiting their applicability in real-time operation. In contrast to the blind estimation approaches, the second strategy is to estimate the 2-D phase error in a semi-blind manner. The main trait of such schemes is to estimate the phase error in a reduced-dimension parameter space by incorporating the *a priori* knowledge of the error structures. Typically, by exploiting the inherent structure of the 2-D phase error, the estimation of the 2-D phase error can be reduced to a 1-D problem. For example, some preliminary knowledge-aided 2-D autofocus methods were proposed in [5-8], by examining the 2-D phase error structure for certain selected image formation algorithms.. These approaches simplify the 2-D phase error into a combination of the APE and the residual RCM. One of the two types of errors is estimated and, based on such results, the other one is computed by exploiting the analytical relationship between these two types of errors. However, after image formation, the 2-D phase error may contain not only the APE and residual RCM, but also high-order terms of the range frequency. Therefore, the image refocused by these algorithms may still suffer from performance degradation when the high-order terms are innegligible. Fig. 5 shows such a data example, where Figs. 5 (a) and 5 (b) are, respectively, the range compressed and full compressed images obtained by using the polar format algorithm (PFA). In Fig. 5 (a) we observe not only the residual RCM but also the range defocus effect, especially at the edges of the aperture.

To obtain an accurate estimate of the residual 2-D phase error using the semi-blind approach, it is a prerequisite to have the *a priori* knowledge about the accurate phase error structure. In this paper, we investigate the inherent structure of the residual 2-D phase error in the framework of PFA. Then, by incorporating the derived inherent phase error structure, we propose a robust 2-D autofocus algorithm which obtained 2-D phase error by estimating phase errors only in a single dimension, i.e., either APE or the residual RCM. The proposed approach provides clear advantages over conventional blind autofocus algorithms in terms of both computational efficiency and estimation accuracy.

This paper is organized as following. In Section II, the inherent structure of the residual 2-D phase errors is analyzed in the PFA framework. Based on this prior knowledge of phase errors, an accurate 2-D autofocus method is proposed in Section III. In Section IV, experimental results are presented to demonstrate the effectiveness of the proposed autofocus approach. Section V presents concluding remarks.

## II. Inherent Structure of 2-D Phase Error in PFA Imagery

### A. *Phase Error in Phase History Domain*

The imaging geometry of a spotlight-mode SAR system is shown in Fig.1. Without loss of generality, we assume that the radar operates in a squint mode with squint angle $\theta_c$. The coordinate of a generic stationary target in the illuminated scene in the $XOY$ plane is $(x_p, y_p)$. Let $t$ represent the slow time. The distance between the antenna phase center (APC) and the scene center (Point O in Fig. 1) is $r_c \equiv r_c(t)$, which, along with the instantaneous azimuth angle $\theta \equiv \theta(t)$ and the incident angle $\varphi \equiv \varphi(t)$, determines the instantaneous coordinate $(x_a, y_a, z_a) \equiv [x_a(t), y_a(t), z_a(t)]$ of the APC. At the aperture center ($t=0$), the incident angle is denoted as $\varphi_{ref}$ and the azimuth angle $\theta = 0$.

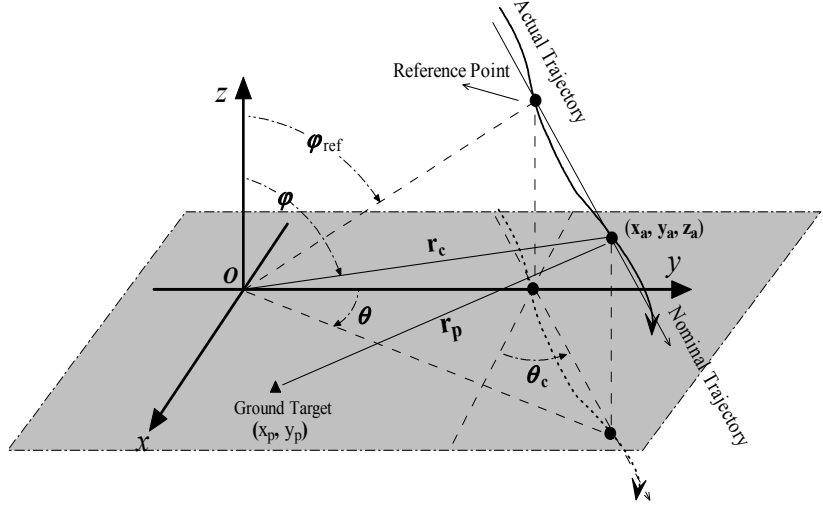

Fig.1.  Spotlight SAR data collection geometry.

To proceed with the PFA, the radar echoes must be converted into the range-frequency domain. The conversion method depends on the type of operating modulations. If a linear frequency modulated (LFM) signal is transmitted, the deramp technique effectively transforms the echoes into the range-frequency domain at the receiver and reduces the required A/D conversion rate. Alternatively, the conversion can be achieved by performing the Fourier transform to the sampled range signal. This approach is applicable to all modulation types. Without loss of generality, our discussion in the sequel will be based on the Fourier transform based approach, whereas extension to the deramp method is straightforward.

After matched filtering and motion compensation with respect to the scene center, the 2-D echo signal can be expressed as

$$S(t, f_r) = A \cdot \exp\left\{j\frac{4\pi}{c}(f_c + f_r) r_\Delta(t)\right\}, \tag{1}$$

where $c$ is the speed of propagation, $f_c$ is the radar center frequency, $f_r$ is the range frequency, $A$ includes the nonessential factors of the transmitted pulse envelop and azimuth antenna pattern, and

$$r_\Delta(t) = r_c(t) - r_p(t) \tag{2}$$

is the differential range, with $r_p(t)$ denoting the instantaneous distance between the APC and the target located at $(x_p, y_p)$.

In the formulation of PFA, the differential range is divided into two terms, i.e., $r_\Delta(t) = r_b(t) + r_e(t)$, where

$$r_b(t) = \sin\varphi\left(x_p \sin\theta + y_p \cos\theta\right) \quad (3)$$

is the basic imaging term and $r_e(t)$ is the error term.

The 2-D signal in (1) can be rewritten as

$$S(t, f_r) = A \cdot \exp\left\{j\frac{4\pi}{c}(f_c + f_r)\left[\sin\varphi\left(x_p \sin\theta + y_p \cos\theta\right) + r_e(t)\right]\right\}. \quad (4)$$

From (4), we can see that the 2-D phase error in the phase history domain can be decomposed into two terms, expressed as

$$\Phi_e(t, f_r) = \frac{4\pi}{c} f_c r_e(t) + \frac{4\pi}{c} r_e(t) f_r. \quad (5)$$

The first term is the APE which is independent of the range frequency, whereas the second term is the residual RCM which is linear to the range frequency. It is clear that, because both terms are proportional to $r_e(t)$, there exists a linear relationship between the APE and residual RCM.

Since most autofocus methods work as postprocessing techniques, we are more concerned with the residual 2-D phase error after image formation processing. In the following, we derive the model of residual 2-D phase error after image formation processing in the framework of PFA.

### B. Phase Error after Polar Reformatting

In this subsection, we analyze the residual 2-D phase error in PFA imagery from two different perspectives. The results will form the basis for our proposed 2-D autofocus method to be discussed in the next section.

**Derivation Method I**

In [19], we provide a new interpretation of polar reformatting, where the range resampling is considered as a range frequency scaling transformation, and the azimuth resampling is interpreted as a combination of RCM linearization and the Keystone transform (KT).

The range frequency scaling transform has a scaling factor of $\delta_r = \sin\varphi_{ref} / (\sin\varphi \cos\theta)$ and an offset of $f_c(\delta_r - 1)$. Therefore, after range resampling, the signal in (4) becomes

$$S_R(t, f_r) = S\left[t, \delta_r f_r + f_c(\delta_r - 1)\right] \\ = A \cdot \exp\left\{j \frac{4\pi(f_c + f_r)\sin\varphi_{ref}}{c}\left[x_p \tan\theta + y_p + \varepsilon(t)\right]\right\}, \tag{6}$$

where $\varepsilon(t) = r_e(t) / (\sin\varphi \cos\theta)$.

The second step of PFA is azimuth resampling. We divide it into two cascaded resampling procedures, i.e., RCM linearization and KT. RCM linearization is a azimuth time transformation, which is independent of the range frequency, to linearize $\tan\theta$. Mathematically, this procedure can be implemented by performing a change-of-variable on the azimuth time, denoted as $t \to \vartheta_a(t)$. Therefore, after RCM linearization, the signal in (6) becomes

$$S_{A1}(t, f_r) = S_R\left[\vartheta_a(t), f_r\right] = A \cdot \exp\left\{j \frac{4\pi(f_c + f_r)\sin\varphi_{ref}}{c}\left[x_p \Omega t + y_p + \eta(t)\right]\right\}, \tag{7}$$

where $\Omega$ is a constant determined by the azimuth resampling process and $\eta(t) = \varepsilon\left[\vartheta_a(t)\right]$.

The final step of polar reformatting is to perform the KT on (7), which results in

$$S_{A2}(t, f_r) = S_{A1}\left[\frac{f_c}{f_c + f_r}t, f_r\right] \\ = A \cdot \exp\left\{j \frac{4\pi \sin\varphi_{ref}}{c}\left[f_c \Omega t x_p + (f_c + f_r)y_p + (f_c + f_r)\eta\left(\frac{f_c}{f_c + f_r}t\right)\right]\right\}. \tag{8}$$

For notational convenience, we define the spatial frequency in azimuth and range, respectively expressed as

$$\begin{aligned} X &= \frac{4\pi \sin\varphi_{ref} f_c \Omega}{c} t \\ Y &= \frac{4\pi \sin\varphi_{ref}}{c}(f_c + f_r) \end{aligned}, \tag{9}$$

where $Y$ has an offset $Y_0 = \dfrac{4\pi \sin \varphi_{ref}}{c} f_c$.

Therefore, in the spatial frequency domain, the signal depicted in (8) can be expressed as

$$S_{A2}(X,\ Y) = A \cdot \exp\left\{j\left[x_p X + y_p Y + Y\eta\left(\dfrac{X}{\Omega Y}\right)\right]\right\}. \tag{10}$$

To simplify the notation, we can define $\xi(X) = \eta\left(\dfrac{X}{\Omega}\right)$. Then (10) becomes

$$S_{A2}(X,\ Y) = A \cdot \exp\left\{j\left[x_p X + y_p Y + Y\xi\left(\dfrac{X}{Y}\right)\right]\right\}. \tag{11}$$

Now it is clear that, after polar reformatting, the residual 2-D phase error in the spatial frequency domain is

$$\Phi_e(X,\ Y) = Y\xi\left(\dfrac{X}{Y}\right). \tag{12}$$

**Derivation Method II**

In the following we examine the residual 2-D phase error after polar reformatting from a different perspective. From [20, 21], we know that there exists a Fourier transform relationship between the terrain reflectivity and the collected data in the phase history domain. That is, the collected data in the phase history domain is 2-D samples in spatial frequency domain of the terrain reflectivity. Each demodulated pulse at a particular azimuth time $t$ along the synthetic aperture yields one line of frequency domain data (actually, only one segment is available due to limited transmitted bandwidth $B$). The line passes through the origin of the coordinate systems and its position is determined by the specific data collection geometry. Note that the azimuth and incident angle of the line in the spatial frequency domain respectively coincide with those in the data collection geometry (see Fig. 2). Therefore, a finite set of echo pulses can provide a polar raster samples in spatial frequency domain of terrain reflectivity.

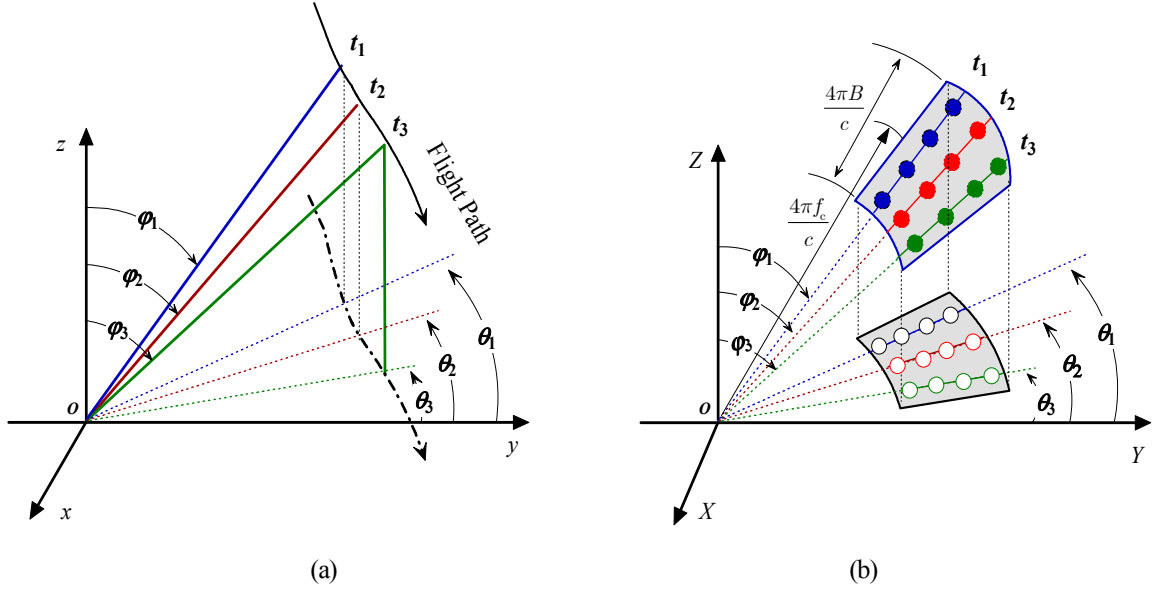

(a)                  (b)

Fig. 2. The relationship between (a) data collection geometry and (b) its corresponding sample positions in spatial frequency domain.

To clearly show these polar format samples, we rewrite (4) as

$$S(\theta, \varphi, K_r) = A \cdot \exp\left\{jK_r\left[\sin\varphi\left(x_p\sin\theta + y_p\cos\theta\right) + r_e(t)\right]\right\}, \tag{13}$$

where $K_r = 4\pi(f_c + f_r)/c$, which along with the azimuth angle $\theta$ and incident angle $\varphi$ determines the sample positions in the spatial frequency domain.

To form the image, only 2-D discrete Fourier transform (DFT) is required. To implement efficient Fourier transform, the fast Fourier transform (FFT) is often a popular choice. Because the 2-D FFT requires uniformly spaced samples on a rectangular grid, the acquired polar samples of phase history must be resampled to a rectangular grid. This is exactly what is done in PFA. After polar reformatting, polar coordinates $(K_r, \theta, \varphi)$ are projected and converted into a Cartesian grid in the ground plane with an azimuth spatial frequency $X$ and a range spatial frequency $Y$.

Considering the following relationship

$$\begin{aligned} X &= K_r\sin\varphi\sin\theta \\ Y &= K_r\sin\varphi\cos\theta \end{aligned}, \tag{14}$$

the signal after polar reformatting can be expressed as

$$S(X, Y) = A \cdot \exp\left\{j\left[x_p X + y_p Y + \sqrt{X^2 + Y^2}\zeta(t)\right]\right\}, \tag{15}$$

where $\zeta(t) = \dfrac{r_e(t)}{\sin\varphi}$.

In (15), $\zeta(t)$ is still a function of azimuth time $t$. In order to analyze the phase error, we desire to express it as a function of $X$ and $Y$. It is clear from Fig. 2 that there is a one-to-one correspondence relationship between the azimuth time $t$ and the ratio $\dfrac{X}{Y}$. By denoting this mapping relationship using the following function

$$t = \kappa\left(\frac{X}{Y}\right), \tag{16}$$

the signal in (15) can be rewritten as

$$S(X, Y) = A \cdot \exp\left\{j\left[x_p X + y_p Y + \sqrt{X^2 + Y^2}\,\mu\left(\frac{X}{Y}\right)\right]\right\}, \tag{17}$$

where $\mu\left(\dfrac{X}{Y}\right) = \zeta\left[\kappa\left(\dfrac{X}{Y}\right)\right]$.

From (17), it become clear that the residual 2-D phase error after polar reformatting is

$$\Phi_e(X, Y) = \sqrt{X^2 + Y^2}\,\mu\left(\frac{X}{Y}\right) = Y\xi\left(\frac{X}{Y}\right), \tag{18}$$

where $\xi\left(\dfrac{X}{Y}\right) = \sqrt{1 + \left(\dfrac{X}{Y}\right)^2}\,\mu\left(\dfrac{X}{Y}\right)$.

### C. Priori Knowledge on Phase Error Structure

In the previous subsection, we have derived the residual 2-D phase error model after polar reformatting from two different perspectives. Different from the one in phase history domain, residual 2-D phase error in the spatial frequency domain includes not only APE and RCM, but also high-order terms in range frequency. To see this effect clearly, we perform a Taylor expansion on (12) or (18) with respect to the range frequency evaluated at $Y_0$

$$\Phi_e(X, Y) = \phi_0(X) + \phi_1(X)(Y - Y_0) + \phi_2(X)(Y - Y_0)^2 + \cdots , \qquad (19)$$

where

$$\begin{aligned} \phi_0(X) &= Y_0 \xi\left(\frac{X}{Y_0}\right) \\ \phi_1(X) &= \xi\left(\frac{X}{Y_0}\right) - \frac{X}{Y_0} \xi'\left(\frac{X}{Y_0}\right), \\ \phi_2(X) &= X^2 \xi''\left(\frac{X}{Y_0}\right) / (2Y_0^3) \end{aligned} \qquad (20)$$

where $\xi'(X/Y_0)$ and $\xi''(X/Y_0)$ respectively denote the first and second derivatives of $\xi(X/Y_0)$. In (19), the $\phi_0(X)$ term is the APE, the $\phi_1(X)$ term corresponds to the residual RCM, and the $\phi_2(X)$ term and other high-order terms are related to the range defocus.

Consulting (20) and (12) or (18), we can find the analytical relationship between the 2-D phase error and the APE as

$$\Phi_e(X, Y) = \frac{Y}{Y_0} \phi_0\left(\frac{Y_0}{Y} X\right), \qquad (21)$$

and the relationship between the 2-D phase error and the residual RCM as

$$\Phi_e(X, Y) = -XY \int \frac{\phi_1\left(\frac{Y_0}{Y} X\right)}{X^2} dX + CX . \qquad (22)$$

In (22), $C$ is an unknown constant which cannot be determined from the residual RCM. Fortunately, its corresponding term is linear to the spatial frequency $X$. Because a linear phase error only introduces an image domain shift, but not affect the focus quality, such linear phase errors can be ignored in the autofocus process which is primarily concerned with the focus quality.

### III. Knowledge-aided Two-dimensional Autofocus for PFA Imagery

#### A. Necessity of 2-D Autofocus

Before applying the 2-D autofocus, we would like to clarify when the use of 2-D autofocus is needed.

From the discussion in the previous section, we know that the residual 2-D phase error in PFA imagery includes three parts, namely, the APE, the residual RCM and the range defocus terms due to high-order terms in the range frequency. In the sequel, the considerations of the range defocus terms is limited to the quadric term, but extension to a general case is straightforward.

Note that 1-D autofocus only estimates and compensates for the APE, and ignores the effect of the residual RCM and the range defocus terms. Such treatment is often sufficient for mid-resolution SAR systems. However, for high-resolution SAR, the dwell time is increased and the range resolution cell size is reduced. As such, the effect of the residual RCM and range defocus terms become more significant. Therefore, for ultra high resolution SAR systems, especially for the systems without motion sensor or the motion sensor has poor accuracy, 2-D autofocus will become more and more popular.

To better understand the condition where 2-D becomes necessary, we consider a simple example where the APE is quadratic, given as

$$\phi_0(X) = aX^2, \qquad (23)$$

where $a$ is a constant coefficient.

From (20), it is easy to obtain the coefficients of the corresponding residual RCM and the range defocus term as

$$\phi_1(X) = \frac{1}{Y_0}\left[\phi_0(X) - X\phi_0'(X)\right] = -\frac{aX^2}{Y_0}, \qquad (24)$$

$$\phi_2(X) = \frac{X^2}{2Y_0^2}\phi_0''(X) = \frac{aX^2}{Y_0^2}, \qquad (25)$$

To avoid significant deterioration in the image quality, a reasonable rule of thumb is to limit these uncompensated errors to an acceptable level. For the APE and range defocus term, it is reasonable to use π/4 rad as a limit, whereas a desirable choice for the residual RCM is for it to be smaller than half a range resolution cell. Denote the range and azimuth resolution cells to be $\rho_y$ and $\rho_x$, respectively. Then, these limitations can be analytically expressed as

APE limit: $\qquad \max\{\phi_0(X)\} - \min\{\phi_0(X)\} \leq \dfrac{\pi}{4}, \qquad (26)$

Residual RCM limit: $\max\{\phi_1(X)\} - \min\{\phi_1(X)\} \leq \dfrac{\rho_y}{2}$, (27)

Range defocus limit: $\max\{\phi_2(X)(Y-Y_0)^2\} - \min\{\phi_2(X)(Y-Y_0)^2\} \leq \dfrac{\pi}{4}$. (28)

Inserting (23)-(25) into (26)-(28), respectively, and taking into account the relationship $\max|X| = \dfrac{\pi}{\rho_x}$ and $\max|Y-Y_0| = \dfrac{\pi}{\rho_y}$, we find that coefficient $a$ must satisfy

APE limit: $a \leq \dfrac{\rho_x^2}{4\pi}$, (29)

Residual RCM limit: $a \leq \dfrac{Y_0 \rho_x^2 \rho_y}{2\pi^2}$, (30)

Range defocus limit: $a \leq \dfrac{Y_0^2 \rho_x^2 \rho_y^2}{4\pi^3}$. (31)

From the above inequalities, it is clear that all the three limits are resolution-dependent. Specifically, the finer the resolution, the more stringent these limits will be. To intuitively show this relationship and give a comparison among the three limits, Fig. 3 presents a graphic illustration, where the radar is assumed to operates in the X-band with an equal resolution both in range and azimuth. In this figure, the parameter space is divided into four areas: (a) In the area below the blue line, the parameters satisfy all the three limits and, therefore, no autofocus is required; (b) In the area between the blue and red lines, the APE limit cannot be satisfied, but the other two conditions are met. In this case, only 1-D autofocus is required; (c) In the area between the red and green lines, the range defocus effect is still negligible, but the APE and the residual RCM cannot be ignored. Therefore, 2-D autofocus becomes necessary; (d) In the area above the green line, all the three limits are violated. in this case, a more accurate 2-D autofocus is required. For some typical values of $a$ in practical systems, e.g., $a = 0.1\ (-10\text{dB})$, 2-D autofocus becomes necessary when the resolution becomes better than 0.1m.

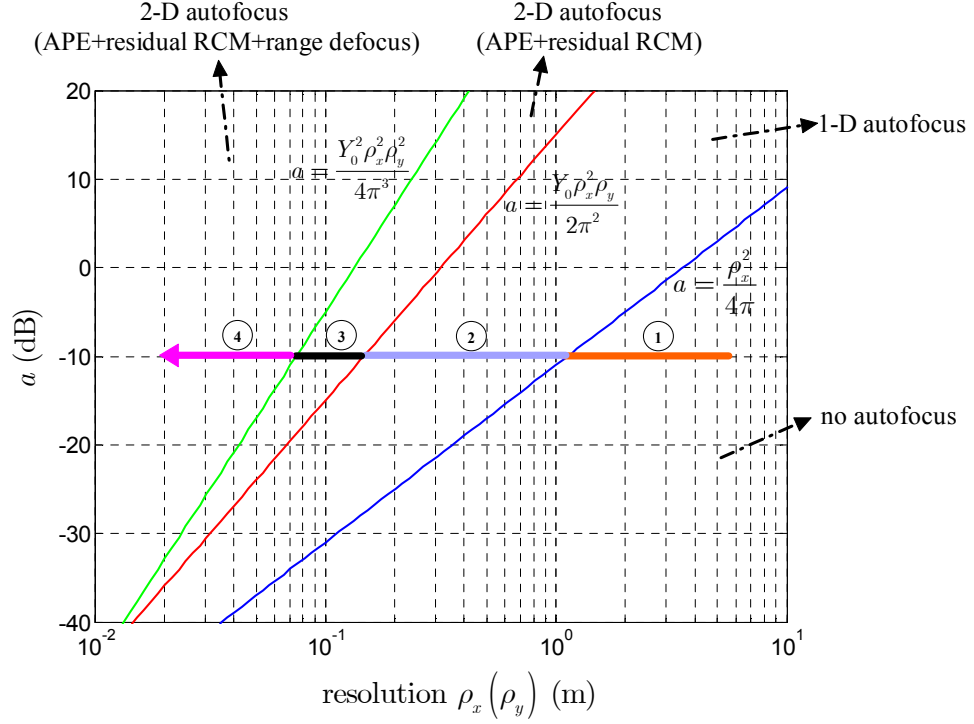

Fig.3. Graphic illustration of the three limits.

## B. *Proposed Method*

In (21) and (22), we obtained the inherent information on the residual 2-D phase error structure, which shows that the 2-D phase error can be completely determined by either the APE or the residual RCM. This means that the estimation of 2-D phase error can be reduced to a 1-D estimation problem by exploiting such *a prior* knowledge about the inherent structure. Typically, it is done by first estimating either the APE or the residual RCM, and then map this 1-D error into a 2-D phase error using (21) or (22).

The APE estimation can typically be implemented using a conventional 1-D autofocus algorithm. But a necessary modification will be required when residual RCM exceeds range resolution cell, since this is not taken into account in conventional method. To solve this problem, at least two alternative strategies at hand can provide this capability. The most straightforward way is to perform a preprocessing on the data to reduce the range resolution [6], thereby keeping the residual RCM smaller than the coarse range resolution cell. After this preprocessing, the APE can be estimated by conventional autofocus techniques such as PGA. The second method is subaperture autofocus which divides a large aperture corresponding to a long coherent processing interval into multiple subapertures. As long as the length of each subaperture is small enough, the residual RCM in each subaperture is negligible and traditional autofocus methods can be used to extract the subaperture phase error. Phase errors from all subapertures are then coherently combined to estimate the overall APE [22]. When the error is not very large, these methods generally yield a good phase estimation accuracy. However, when the error become very large, which is just the case we need to consider in this work, they may suffer serious performance degradation.

For the residual RCM estimation, there is little work reported in literature. One possible solution is to draw lessons from range alignment in inverse SAR (ISAR). Typically, we can measure the range shift directly in the range compressed data by correlating the entire range profile. The estimate accuracy of RCM obtained from these methods may be adequate to align the range profiles within a range resolution cell. Such results, however, are difficult to satisfy the phase compensation requirement when they are mapped to the 2-D phase error. Nevertheless, the advantage of the RCM estimation is that its performance is not affected by the size of error.

To obtain an accurate estimate of 2-D phase error, we combine the above two methods to overcome their respective shortcomings. The overall processing flow of the yielding approach is depicted in Fig. 4. First, we estimate the residual RCM in the range compressed data domain and map it into a 2-D phase error using (22). The estimated 2-D phase error is then applied to the defocused image to produce a coarsely focused one. This compensation is not necessarily to be very accurate, but the correction procedure works well even when the error is very large. After this coarse compensation, most of the 2-D phase error has been corrected. The residual 2-D phase error becomes relative small. In this case, an APE estimate and its mapping to 2-D phase error using (21) can be performed to achieve an accurate estimate of the residual 2-D phase error. Finally, the estimated residual 2-D phase error is applied to the coarsely focused image to produce a fine focused image.

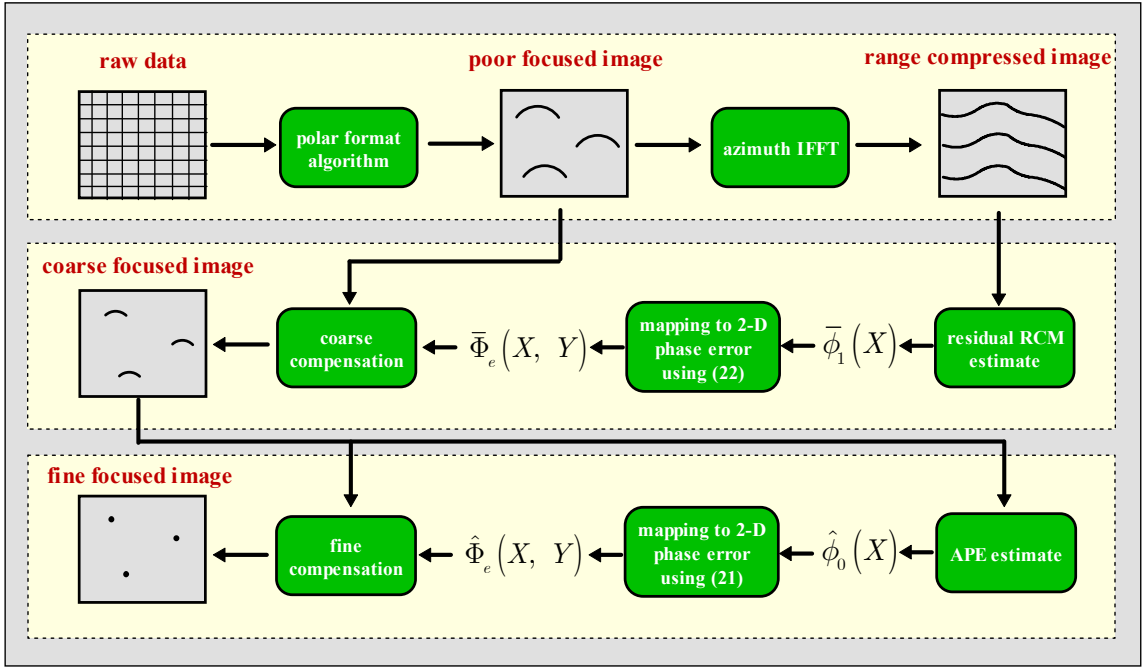

Fig. 4. Flowchart of the proposed 2-D autofocus algorithm.

## C. Some Practical Issues

In this subsection, we address two important issues to be considered in practice. The first issue we are concerned is the estimation error of the proposed method. Since the residual 2-D phase error in our method is computed from the estimated APE and/or residual RCM, the estimation accuracy of the 2-D phase error will depend on that of the APE and the residual RCM. For APE estimation, it is well-known that all the existing methods have an inherent drawback that they cannot estimate the constant and linear phase terms. Without loss of generality, we assume that these two terms are denoted as $\Delta\phi_0(X) = a_0 + a_1 X$, where $a_0$ and $a_1$ are arbitrary constants. According to (21), the corresponding 2-D phase error is expressed as

$$\Delta\Phi_e(X, Y) = \frac{Y}{Y_0} \Delta\phi_0\left(\frac{Y_0}{Y} X\right) = \frac{a_0}{Y_0} Y + a_1 X, \qquad (32)$$

which is linear in $X$ and $Y$ domains. These linear phase error only introduce a displacement in the image domain. Therefore, the inherent bias of the APE estimation will not affect the focus quality in the proposed 2-D autofocus method.

For RCM estimation, there exists a similar problem. Existing range alignment methods only estimate the relative range migration but the constant bias is often ignored. By denoting the constant bias as $\Delta\phi_1(X) = b$, we can obtain the corresponding 2-D phase error according to (22) as

$$\Delta\Phi_e(X, Y) = -XY \int \frac{\Delta\phi_1\left(\frac{Y_0}{Y}X\right)}{X^2} dX = bY. \tag{33}$$

This is also a linear phase term which does not cause image defocus.

The second issue to be addressed is the space-variant characteristics of the phase error. In the PFA imagery, there are two main sources of phase error. One is the range error caused by SAR sensor motion or turbulent media. The other is due to algorithm approximation. Both phase errors are in essence space-variant because they are different for different scatterers in the image. A common approach to space-variant autofocus is to partition the large scene into multiple smaller subscenes such that the error in each subscene can be considered space-invariant. The subimage of each subscene is then focused independently using space-invariant autofocus algorithms, and all the refocused subimages are mosaicked together to yield a focused full-scene image. This strategy can also be adopted in our 2-D autofocus method. It is noteworthy that the phase error caused by algorithm approximation is deterministic. Of course it can be estimated and corrected along with the other errors by the autofocus procedure. However, to reduce the heavy burden on autofocus, we can also correct for it before autofocus is applied. In PFA imagery, the main phase error caused by algorithm approximation is the range curvature error. This error can be compensated for by a space-variant post-processing procedure[23, 24].

## IV. Experimental Results

Raw data collected by an airborne spotlight SAR system is applied to demonstrate the effectiveness of the proposed autofocus methods. The radar operates at the X-band with a signal bandwidth of 1.3 GHz, which correspond to a theoretical range resolution of about 0.12m. During the data collection, the radar platform flew at a height of about 5000 m with an average velocity of 120 m/s. The slant range from the radar to the scene center is about 8 km. The synthetic aperture length we processed in this experiment is 1760 m, which produces a nominal azimuth resolution of 0.07 m. A motion sensing system consisting of a GPS and an IMU is equipped to provide a measurement of the radar's motion. However, in the image formation procedure, we did not exploit this sensor data, but assume that the radar platform flew at a constant velocity. The reasons are twofold. First, the accuracy of the data provided by this sensor is relatively poor. It can greatly improve the focus quality but still can't provide a satisfactory result. Second, the use of the uncompensated data for image formation enables us to better illustrate the refocus performance of the proposed algorithm in the presence of large errors.

Figs. 5(a) and 5(b) show the range compressed image (to show residual RCM clearly, magnified local image including strong scatterers is presented) and fully compressed image produced by the PFA processing with no autofocus procedures applied. It is observed that, although deterministic range migration is compensated by polar format transformation, residual RCM is still large enough to exceed range resolution cells. As a matter of fact, the images suffer from not only the residual RCM but also the range defocus effect. Such effect is obvious in the range compressed image as shown in Fig. 5(a).

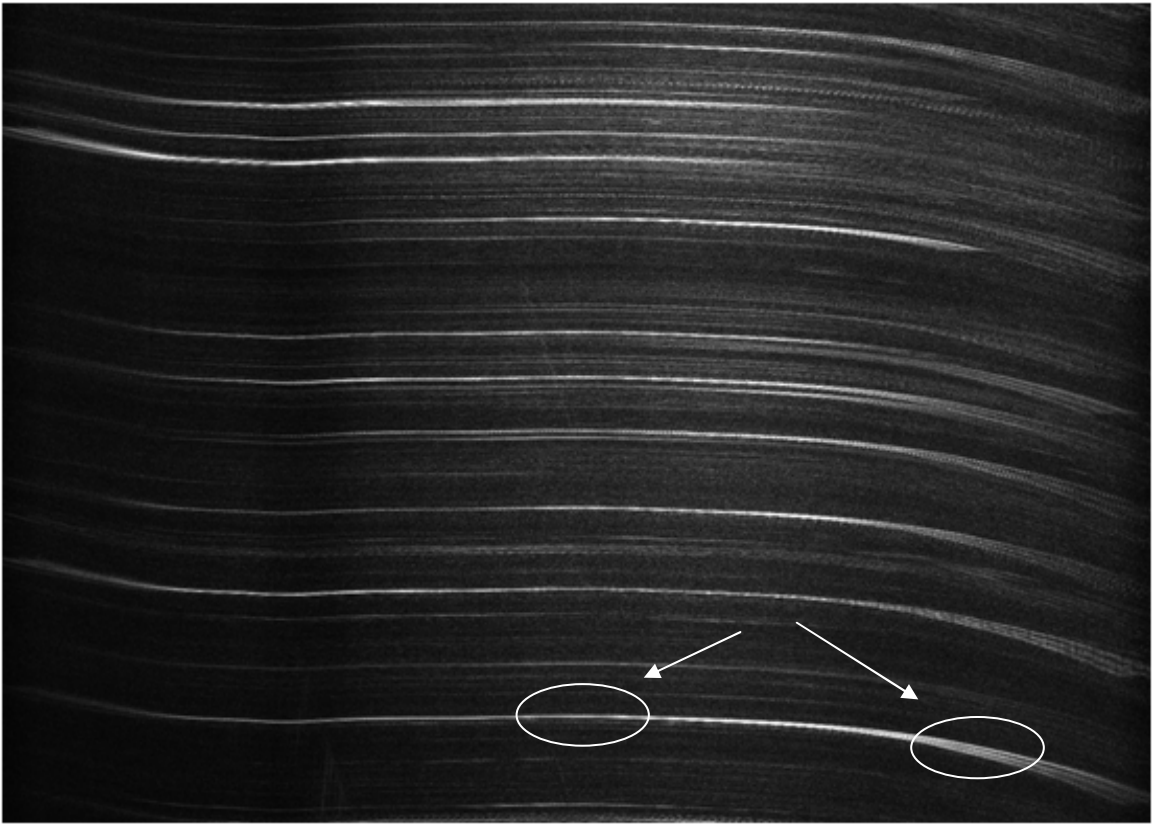

(a)

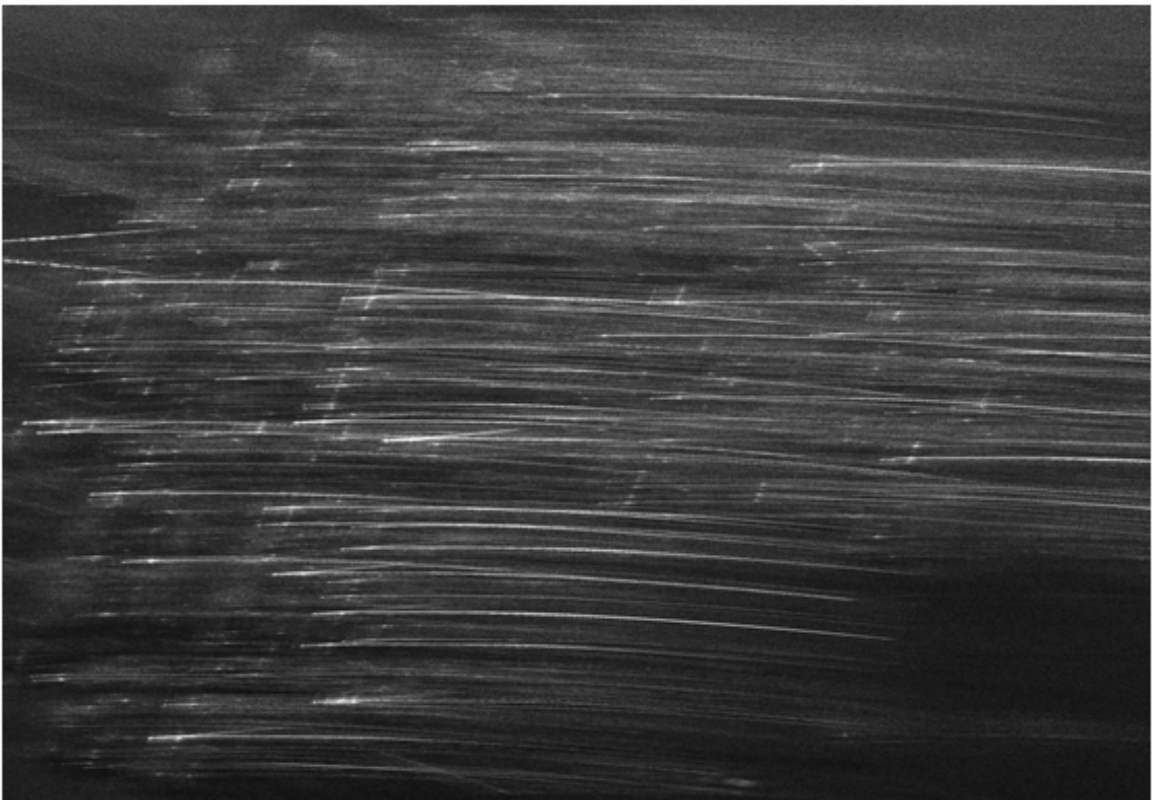

(b)

Fig.5. (a) Range compressed and (b) fully compressed images after PFA processing.

Conventional 1-D autofocus methods only remove the azimuth phase error, while the residual RCM and high-order terms in range frequency are ignored. This simplification is sufficient for SAR systems with a moderate resolution, where the synthetic aperture is relatively small. However, in this experiment where the resolution is very high and no motion sensor data is applied, 1-D autofocus obviously can't meet the accuracy requirement. Fig. 6 shows such an example, where the refocused image is obtained by 1-D autofocus processing using MD-PGA[22]. Although the focus quality is substantially improved, the residual 2-D defocusing effect remains significant.

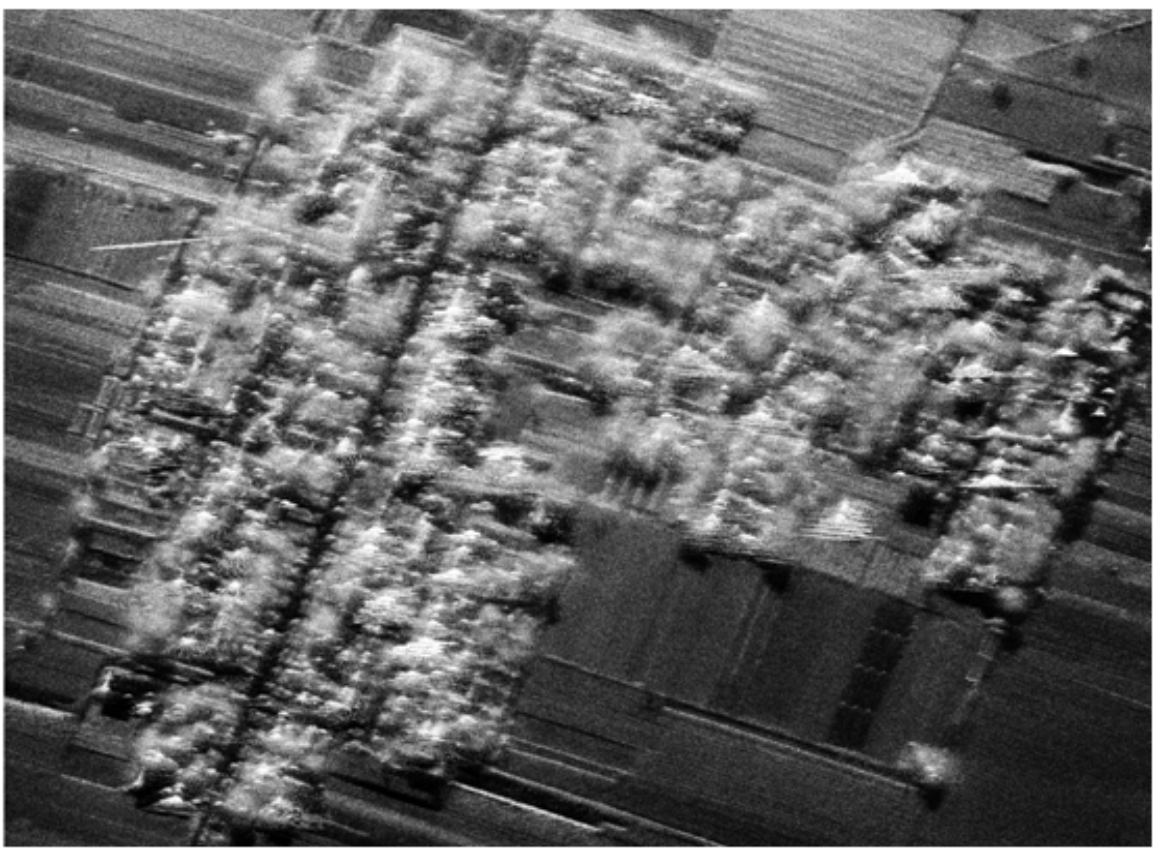

Fig.6. Refocused image by MD-PGA

The 2-D autofocus proposed in [5] is also applied to refocus the PFA image, and the results are shown in Fig. 7. Clearly, the focus quality is greatly improved, since both the APE and the residual RCM are compensated for. However, high-order terms in range frequency are not taken into account in this algorithm, thus yielding substantial residual degradation. From the range compressed image, we can clearly see that the residual RCM has been almost perfectly corrected. However, the range defocus effect still remain.

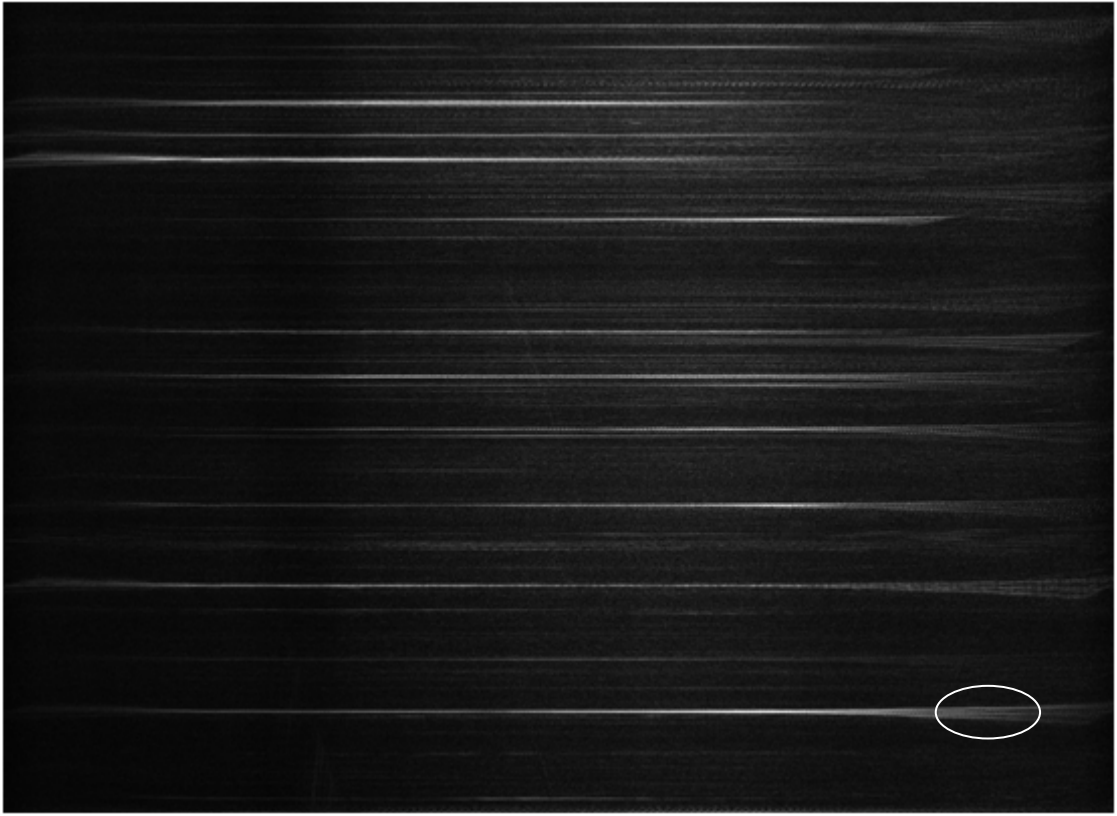

(a)

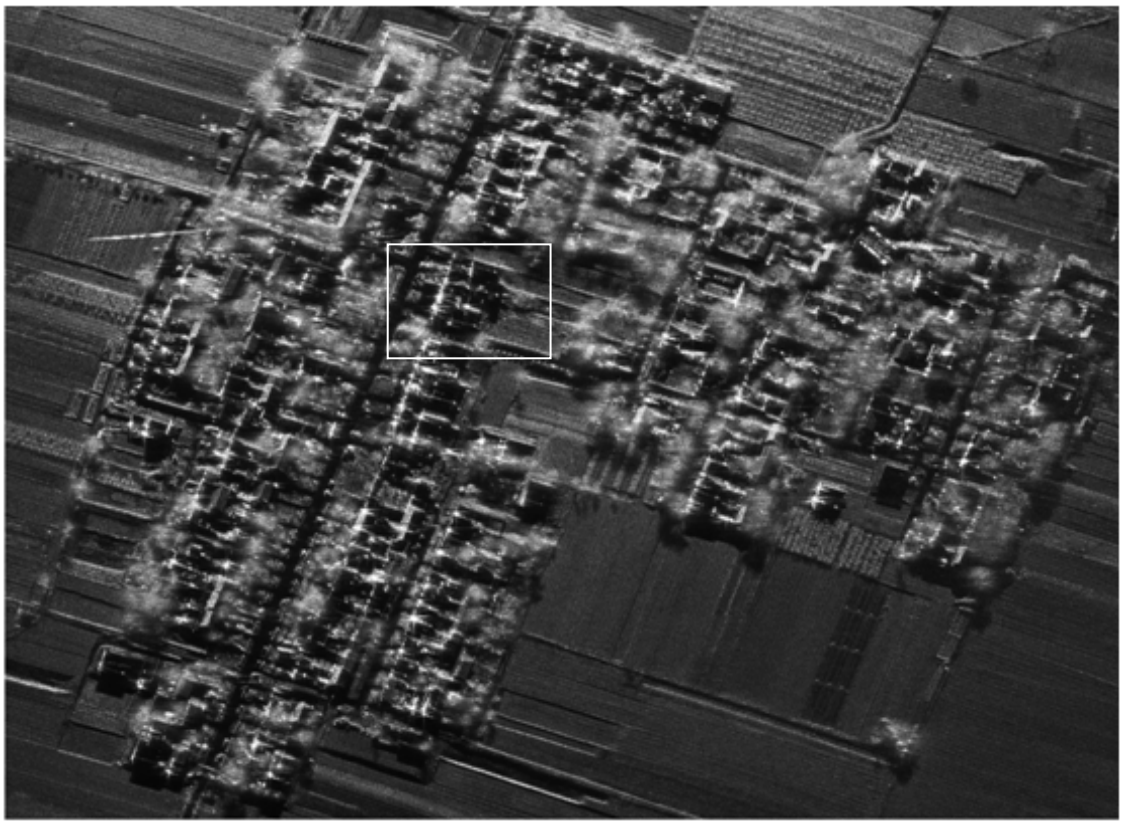

(b)

Fig.7. (a) Range compressed and (b) fully compressed images after 2-D autofocus proposed in [5].

**21 / 27**

Finally, the proposed 2-D autofocus method in this work is applied to refocus the PFA image. In this processing, we use the simplified range tracking method proposed in [17] to estimate the residual RCM, and the MD-PGA proposed in [22] is used to estimate the APE. Fig. 8 shows the processing results, where Fig. 8(a) depicts the range compressed image and Fig. 8(b) the fully compressed image. It is evident from Fig. 8(a) that both the residual RCM and the range defocus effect are eliminated. To better demonstrate the improvement on the focus quality, Fig. 9(a) and 9(b) show the magnified local scenes of Figs. 7(b) and 8(b), respectively.

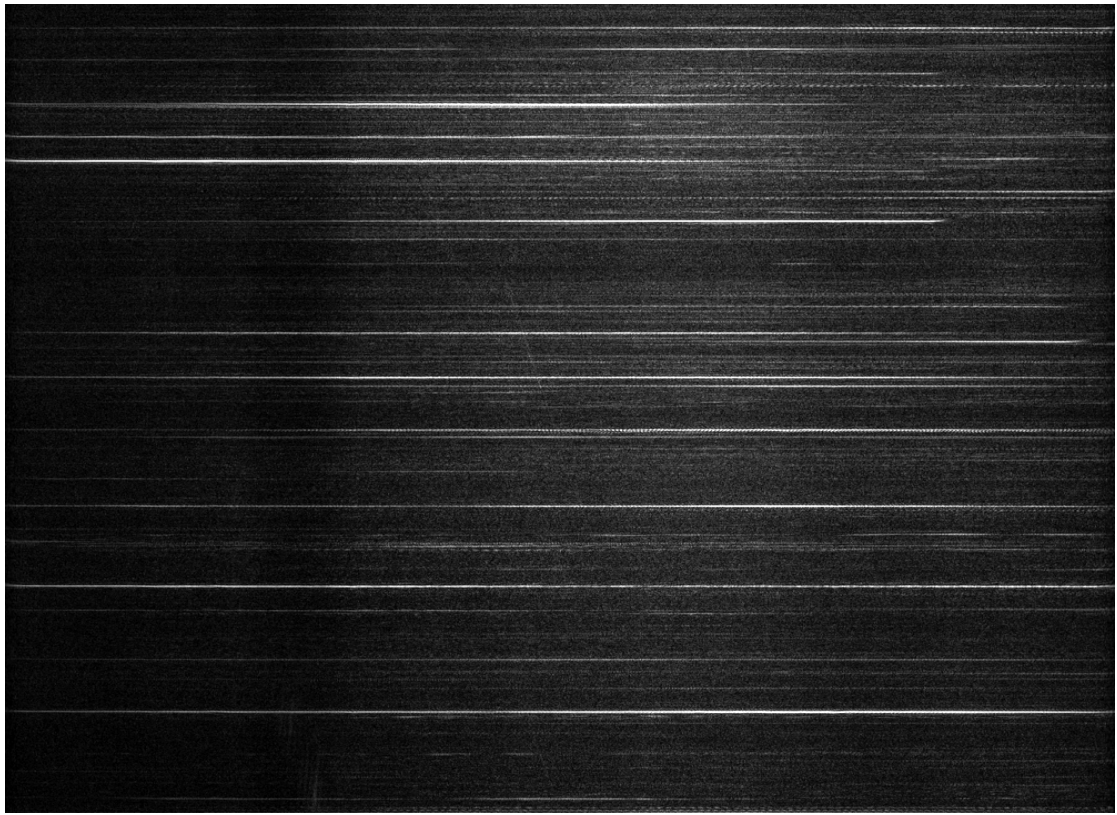

(a)

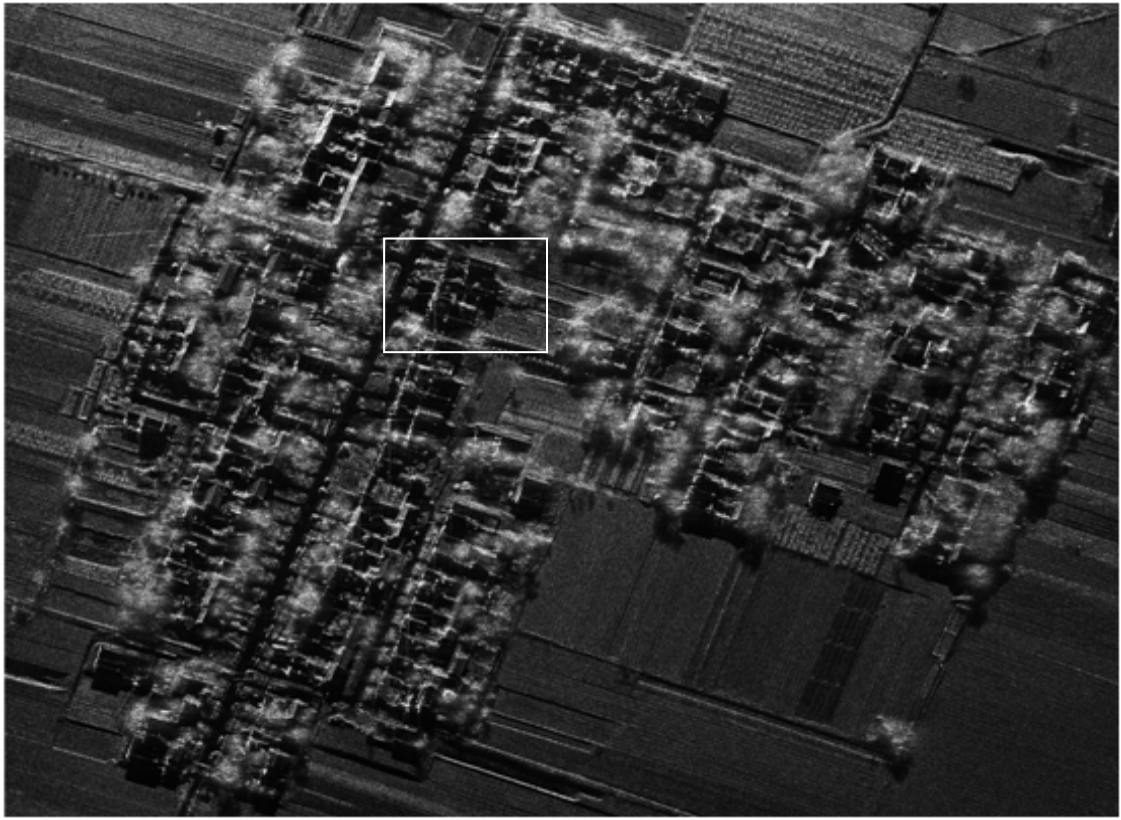

(b)

Fig.8. (a) Range compressed and (b) full compressed images after processing by the proposed method.

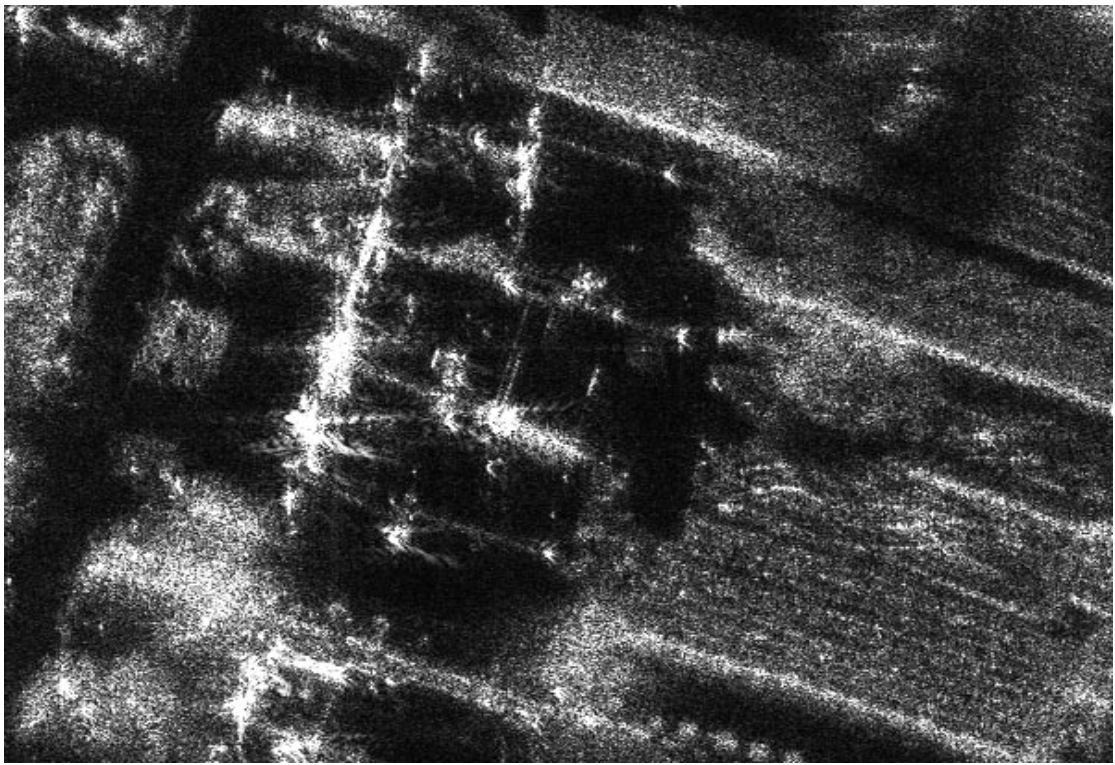

(a)

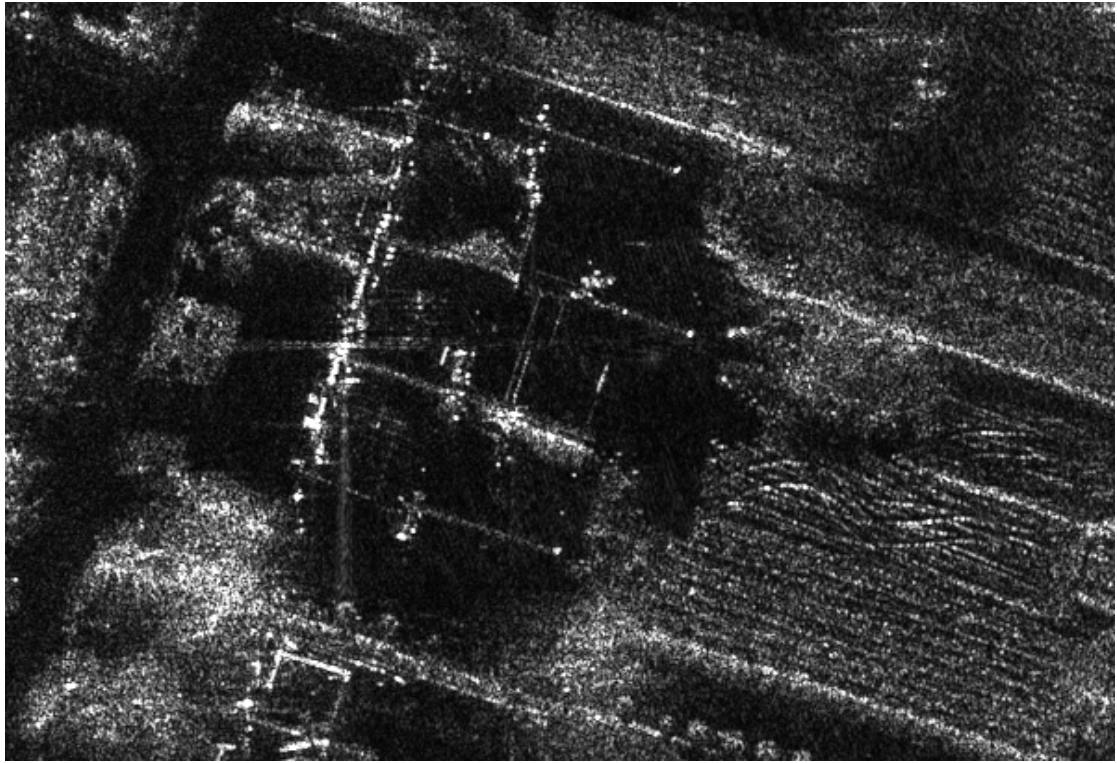

(b)

Fig. 9 Magnified local scene of (a) Fig. 7 and (b) Fig. 8

## V. Conclusions

In this paper, we have analyzed the effect of polar reformatting in PFA on uncompensated 2-D phase error and revealed the analytical structure of residual 2-D phase error in the spatial frequency domain for PFA imagery. By incorporating the derived *a priori* knowledge on the phase error structure, we then proposed a robust 2-D autofocus algorithm which achieves 2-D phase error estimation and correction based only on 1-D phase error estimation. Because the parameter estimation is performed in the reduced-dimension space, the proposed approach offers clear advantages in both computational efficiency and estimation accuracy as compared with conventional blind 2-D autofocus algorithms.

ACKNOWLEDGMENTS

Valuable discussions and collaboration with Daiyin Zhu and Yimin D Zhang are gratefully acknowledged.

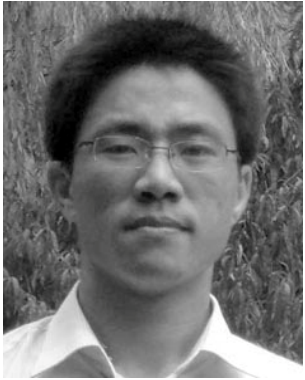

**Xinhua Mao** was born in Lianyuan, China, in 1979. He received the B.S. and Ph.D. degrees from the Nanjing University of Aeronautics and Astronautics (NUAA), Nanjing, China, in 2003 and 2009, respectively, all in electronic engineering.

    Since 2009, he joined the Department of Electronic Engineering, NUAA, where he is now an associate professor. His research interests include radar imaging, and ground moving target indication (GMTI).